\documentclass[twocolumn,tighten]{aastex631}

\usepackage{graphicx}	
\usepackage{amsmath}	
\usepackage{amssymb}	
\usepackage{bm}		
\usepackage{rotating}
\usepackage{longtable}

\def\lea{\mathrel{<\kern-1.0em\lower0.9ex\hbox{$\sim$}}}
\def\gea{\mathrel{>\kern-1.0em\lower0.9ex\hbox{$\sim$}}}

\newcommand{\UWyoming}{\affiliation{Department of Physics and Astronomy, University of Wyoming, Laramie, WY 82071, USA}}
\newcommand{\STScI}{\affiliation{Space Telescope Science Institute, 3700 San Martin Drive, Baltimore, MD 21218, USA}}
\newcommand{\UAntof}{\affiliation{Centro de Astronomía (CITEVA), Universidad de Antofagasta, Avenida Angamos 601, Antofagasta, Chile}}
\newcommand{\NOIRLab}{\affiliation{Gemini Observatory/NSF NOIRLab, 950 N. Cherry Avenue, Tucson, AZ 85719, USA}}
\newcommand{\UToledo}{\affiliation{Ritter Astrophysical Research Center, University of Toledo, Toledo, OH 43606, USA}}
\newcommand{\JHU}{\affiliation{Department of Physics and Astronomy, The Johns Hopkins University, Baltimore, MD 21218 USA}}

\newcommand{\OSU}{\affiliation{Department of Astronomy, The Ohio State University, 140 West 18th Ave., Columbus, OH 43210, USA}}
\newcommand{\MPIA}{\affiliation{Max Planck Institut f\"ur Astronomie, K\"onigstuhl 17, 69117 Heidelberg, Germany}}
\newcommand{\ESO}{\affiliation{European Southern Observatory,  Karl-Schwarzschild Str. 2, 85748 Garching bei Muenchen, Germany}}
\newcommand{\UAlberta}{\affiliation{Department of Physics, University of Alberta, Edmonton, AB T6G 2E1, Canada}}

\newcommand{\Belgium}{\affiliation{Sterrenkundig Observatorium, Universiteit Gent, Krijgslaan 281 S9, B-9000 Gent, Belgium}}
\newcommand{\UHeidelberg}{\affiliation{Astronomisches Rechen-Institut, Zentrum für Astronomie der Universität Heidelberg, Mönchhofstr. 12-14, D-69120 Heidelberg, Germany}}
\newcommand{\Bonn}{\affiliation{Argelander-Institut für Astronomie, Universität Bonn, Auf dem Hügel 71, 53121, Bonn, Germany}}

\newcommand{\ITA}{\affiliation{Institut f\"{u}r Theoretische Astrophysik, Zentrum f\"{u}r Astronomie der Universit\"{a}t Heidelberg,\\ Albert-Ueberle-Strasse 2, 69120 Heidelberg, Germany}}
\newcommand{\COOL}{\affiliation{Cosmic Origins Of Life (COOL) Research DAO, coolresearch.io}}
\newcommand{\IWR}{\affiliation{Universit\"{a}t Heidelberg, Interdisziplin\"{a}res Zentrum f\"{u}r Wissenschaftliches Rechnen, Im Neuenheimer Feld 205, D-69120 Heidelberg, Germany}}
\newcommand{\MPE}{\affiliation{Max-Planck-Institut f\"ur Extraterrestrische Physik (MPE), Giessenbachstr. 1, D-85748 Garching, Germany}}

\newcommand{\OCADU}{\affiliation{OCAD University, Toronto, Ontario, M5T 1W1, Canada}}
\newcommand{\UArizona}{\affiliation{Steward Observatory, University of Arizona, 933 N Cherry Ave,Tucson, AZ 85721, USA}}
\newcommand{\Riverside}{\affiliation{Department of Physics and Astronomy, University of California, Riverside, CA, 92521 USA}}
\newcommand{\Connecticut}{\affiliation{University of Connecticut, Department of Physics, 196A  Auditorium Road, Unit 3046, Storrs, CT, 06269}}
\newcommand{\ARI}{\affiliation{Astrophysics Research Institute, Liverpool John Moores University, 146 Brownlow Hill, Liverpool L3 5RF, UK}}
\newcommand{\LaPlata}{\affiliation{Instituto de Astrofisica de La Plata, CONICET–UNLP,
 Paseo del Bosque S/N, B1900FWA La Plata, Argentina }}
\newcommand{\sorbonne}{\affiliation{Sorbonne {Universit\'e}, LERMA, Observatoire de Paris, PSL university, CNRS, F-75014, Paris, France}}
\newcommand{\CITA}{\affiliation{Canadian Institute for Theoretical Astrophysics (CITA), University of Toronto, 60 St George Street, Toronto, ON M5S 3H8, Canada}}
\newcommand{\MCMASTER}{\affiliation{Department of Physics and Astronomy, McMaster University, 1280 Main Street West, Hamilton, ON L8S 4M1, Canada}}

\shorttitle{Massive Young Star Clusters and New Insights from JWST}
\shortauthors{Whitmore et al.}

\begin{document}

\title{PHANGS-JWST First Results: Massive Young Star Clusters and New Insights from JWST Observations of NGC~1365}


\correspondingauthor{Bradley~C.~Whitmore}
\email{whitmore@stsci.edu}
\author[0000-0002-3784-7032]{Bradley~C.~Whitmore}
\STScI
\author[0000-0003-0085-4623]{Rupali~Chandar}
\UToledo
\author[0000-0002-0579-6613]{M. Jimena Rodríguez}
\UArizona
\LaPlata
\author[0000-0003-0946-6176]{Janice~C.~Lee}
\NOIRLab
\author[0000-0002-6155-7166]{Eric~Emsellem}
\ESO
\author{Matthew~Floyd}
\UToledo
\author{Hwihyun~Kim}
\NOIRLab
\author[0000-0002-8804-0212]{J.~M.~Diederik~Kruijssen}
\COOL
\author[0000-0001-7413-7534]{Angus~Mok}
\OCADU
\author{Mattia~C.~Sormani}
\UHeidelberg
\author[0000-0003-0946-6176]{M\'ed\'eric~Boquien}
\UAntof
\author[0000-0002-5782-9093]{Daniel~A.~Dale}
\UWyoming
\author[0000-0001-5310-467X]{Christopher~M.~Faesi}
\Connecticut
\author[0000-0001-7448-1749]{Kiana~F.~Henny}
\UWyoming
\author{Stephen~Hannon}
\Riverside
\author[0000-0002-8528-7340]{David~A.~Thilker}
\JHU
\author{Richard~L.~White}
\STScI
\author[0000-0003-0410-4504]{Ashley~T.~Barnes}
\Bonn
\author[0000-0003-0166-9745]{F. Bigiel}
\Bonn
\author[0000-0002-5635-5180]{M\'elanie Chevance}
\ITA
\COOL
\author[0000-0001-9656-7682]{Jonathan~D.~Henshaw}
\ARI
\MPIA
\author[0000-0002-0560-3172]{Ralf S.\ Klessen}
\ITA
\IWR
\author[0000-0002-2545-1700]{Adam~K.~Leroy}
\OSU
\author[0000-0001-9773-7479]{Daizhong Liu}
\MPE
\author[0000-0001-6038-9511]{Daniel Maschmann}
\UArizona
\sorbonne
\author[0000-0002-6118-4048]{Sharon~E.~Meidt}
\Belgium
\author[0000-0002-5204-2259]{Erik~Rosolowsky}
\UAlberta
\author[0000-0002-3933-7677]{Eva~Schinnerer}
\MPIA
\author{Jiayi~Sun}
\CITA
\MCMASTER
\author[0000-0002-7365-5791]{Elizabeth~J.~Watkins}
\UHeidelberg
\author[0000-0002-0012-2142]{Thomas~G.~Williams}
\MPIA

\begin{abstract}

A primary new capability of JWST is the ability to penetrate the dust in star forming galaxies to identify and study the properties of 
young star clusters that remain embedded in dust and gas.
In this paper we combine new infrared images taken with JWST with our optical HST images of the star-bursting barred (Seyfert2) spiral galaxy NGC 1365. 
We find that this galaxy has the richest population of massive young clusters of any known galaxy within 30 Mpc, with $\sim$ 30 star clusters that are more massive than $10^{6}~M_{\odot}$ and younger than 10 Myr. Sixteen of these clusters are newly discovered from our JWST observations.  An examination of the optical images reveals that 4 of 30 ($\sim 13$\%)  are so deeply embedded that they cannot be seen in the Hubble I band ($A_V \gea 10$~mag), and that 11 of 30 ($\sim37$\%) are missing in the HST B  band, 
so age and mass estimates from optical measurements alone are challenging.  These numbers suggest that massive clusters in NGC~1365 remain completely obscured in the visible for $\sim1.3\pm0.7$~Myr, and are either completely or partially obscured for $\sim3.7\pm1.1$~Myr. 
We also use the JWST observations to gain new insights into
the triggering of star cluster formation by the collision of gas and dust streamers with gas and dust in the bar.  The JWST images reveal previously unknown structures (e.g., bridges and overshoot regions from stars that form in the bar) that help us better understand the orbital dynamics of barred galaxies and associated star-forming rings.
Finally, we note that the excellent spatial  resolution of the NIRCAM F200W filter provides a  better way to separate barely resolved compact clusters from individual stars based on their sizes. 


\end{abstract}


\section{Introduction and Motivation} \label{sec:intro}

The discovery that massive young Super Star Clusters (SSCs) are forming in merging and starbursting galaxies was one of the earliest results from the Hubble Space Telescope (e.g., \citealt{holtzman92},
\citealt{whitmore93}, 
\citealt{O'connell94}, \citealt{barth95}).  
This discovery provided a direct observational window into the conditions needed to form massive clusters, some of which
will likely evolve into globular clusters (e.g., \citealt{ashman92}, \citealt{elmegreen97},
\citealt{kruijssen2015}).
Many of these SSCs form in dusty regions, including luminous infrared galaxies (e.g.,
\citealt{Adamo20}, \citealt{Linden21}) 
with radio observations identifying some of the youngest, most embedded ones (e.g., \citealt{sandqvist95}, \citealt{whitmore02}, \citealt{Johnson03}, \citealt{He22}).

The dusty, barred spiral galaxy NGC~1365, at a distance of 19.57 $\pm$ 0.78 Mpc (\citealt{gagandeep21}), has one of the highest rates of star formation (16.9 M$_{\odot}~\mbox{yr}^{-1}$) of any galaxy within 20~Mpc (the next  highest in the PHANGS-HST sample of 38 galaxies is 7.6 M$_{\odot}~\mbox{yr}^{-1}$ - \citealt{lee22}), and is known to have recently (past $\approx10$~Myr) formed very massive ($\gea 10^6~M_{\odot}$) clusters \citep[e.g.,][]{kristen97,galliano08}.  

The main goal of this paper is to use
new infrared imaging of NGC~1365, taken with the recently launched 
JWST, plus our optical HST images to collect a complete census of these very young massive clusters, with a focus on discovering clusters missed by previous optical studies because they are enshrouded in dust.

The rest of this paper is organized as follows. Section
\ref{sec:data} 
describes the observations and photometric reductions. Section
\ref{sec:reg_1} examines various objects in Region~1, the northern half of the central 
star-forming ring,
and compiles a census of massive star clusters. 
Section \ref{sec:reg_2} 
examines some interesting features in Region~2, just outside the central region along the eastern bar. 
Our main findings are summarized in Section \ref{sec:conclusions}.

\section{Observations and Reductions}
\label{sec:data}

The primary datasets used in this paper are from the PHANGS-JWST Treasury Proposal (PI: J.C. Lee, GO-2107), and the PHANGS-HST project (PI: J.C. Lee, GO-15654). 

PHANGS-HST is a Cycle 26 HST Treasury program which obtained imaging for 38 nearby spiral galaxies in the following five filters: F275W (NUV), F336W (U), F438W (B), F555W (V) and F814W (I), with the WFC3 camera (or ACS in cases where archival data was available). See \citet{lee22}\footnote{\url{https://archive.stsci.edu/hlsp/phangs-hst}}  
 for details about PHANGS-HST.

PHANGS-JWST is a Cycle 1 JWST Treasury program
which obtained imaging for the 19 (out of 38) PHANGS-HST galaxies that have IFU maps obtained with VLT/MUSE \citep{emsellem22}. ALMA CO(2-1) observations are  available for a larger set of  90 PHANGS-ALMA \citep{leroy21} galaxies that include all 19 of the PHANGS-JWST subset.
The JWST observations  were made in the NIRCAM filters F200W, F300M, F335M, and F360M; and in the MIRI filters F770W, F1000W, F1130W, and F2100W. See \citet{lee22b} for additional details about the observations and basic reductions.

Aperture photometry was performed using a $0.155\arcsec$ radius aperture with a sky annulus between $0.217\arcsec$ and $0.310\arcsec$, as described in \citet{rodriguez22}.  This subtends a five pixel radius for the  F200W  NIRCAM observations, and includes approximately 75~\% of the light from a point source, but smaller percentages at longer wavelengths (e.g., $\sim40$ \% for the F1000W filter and $\sim10$ \% for F2100W). For this reason, the JWST magnitudes are used in a relative rather than absolute sense throughout this paper.  Magnitudes from Hubble are referred to as NUV, U, B, V, I and those from JWST are referred to by their filter names in order to avoid confusion.
Source selection was performed  based on the F335M image, and the corresponding positions were then measured in the other bands. 
The ABmag system is used throughout.


In the future, aperture corrections will be determined for both stars and clusters. A variety of other observational parameters 
will also improve with time, including image alignment and geometric solutions, photometric zeropoints, and background subtraction, as discussed in \citet{lee22b}.  
For these reasons the data and massive cluster catalog we compile here should be considered preliminary. A more definitive treatment will be provided in future papers.

\section{Region 1: Massive Young Star Clusters}
\label{sec:reg_1}

\begin{figure}
\begin{center}
\includegraphics[width =3.3in , angle= 0]{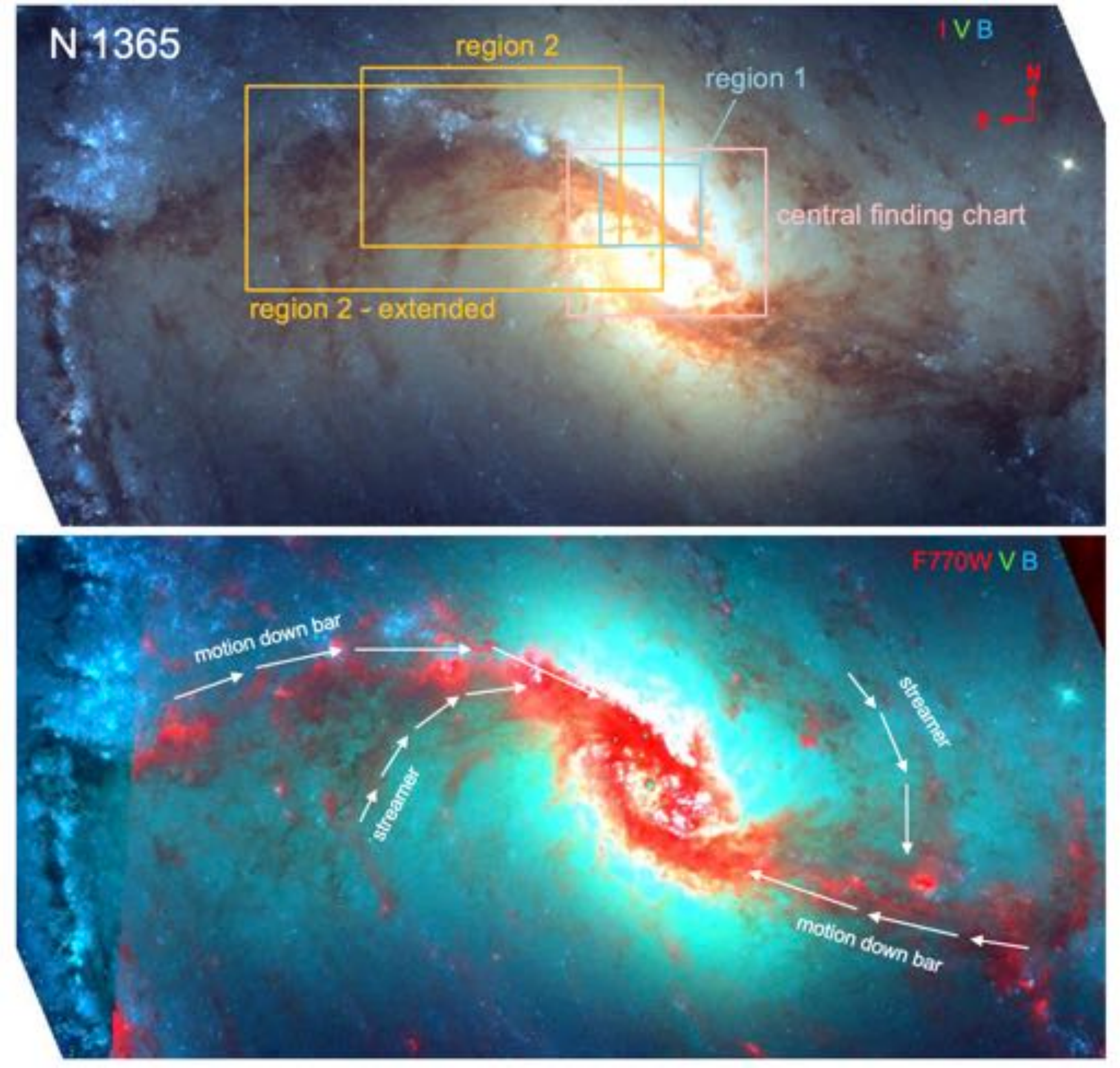}

\end{center}
\caption{
An optical IVB  color image taken by Hubble of NGC~1365 is shown in the upper panel,  and a JWST infrared (F770W) plus Hubble (V and B) image is shown in the bottom panel. The full field of view covers an area of roughly 80$''$ $\times$ 160$''$ (7.5 kpc $\times$ 15.0 kpc). The fields-of-view for regions discussed in the text are shown, as well as the motion along the bar and along two of the  streamers. We note that the extensive star formation in the spiral arm and the end of the Eastern bar on the left edge of the image is outside the F770W field of view.}
\label{fig:full_galaxy}
\end{figure}
\begin{figure*}
\begin{center}
\includegraphics[width =6.5in , angle= 0]{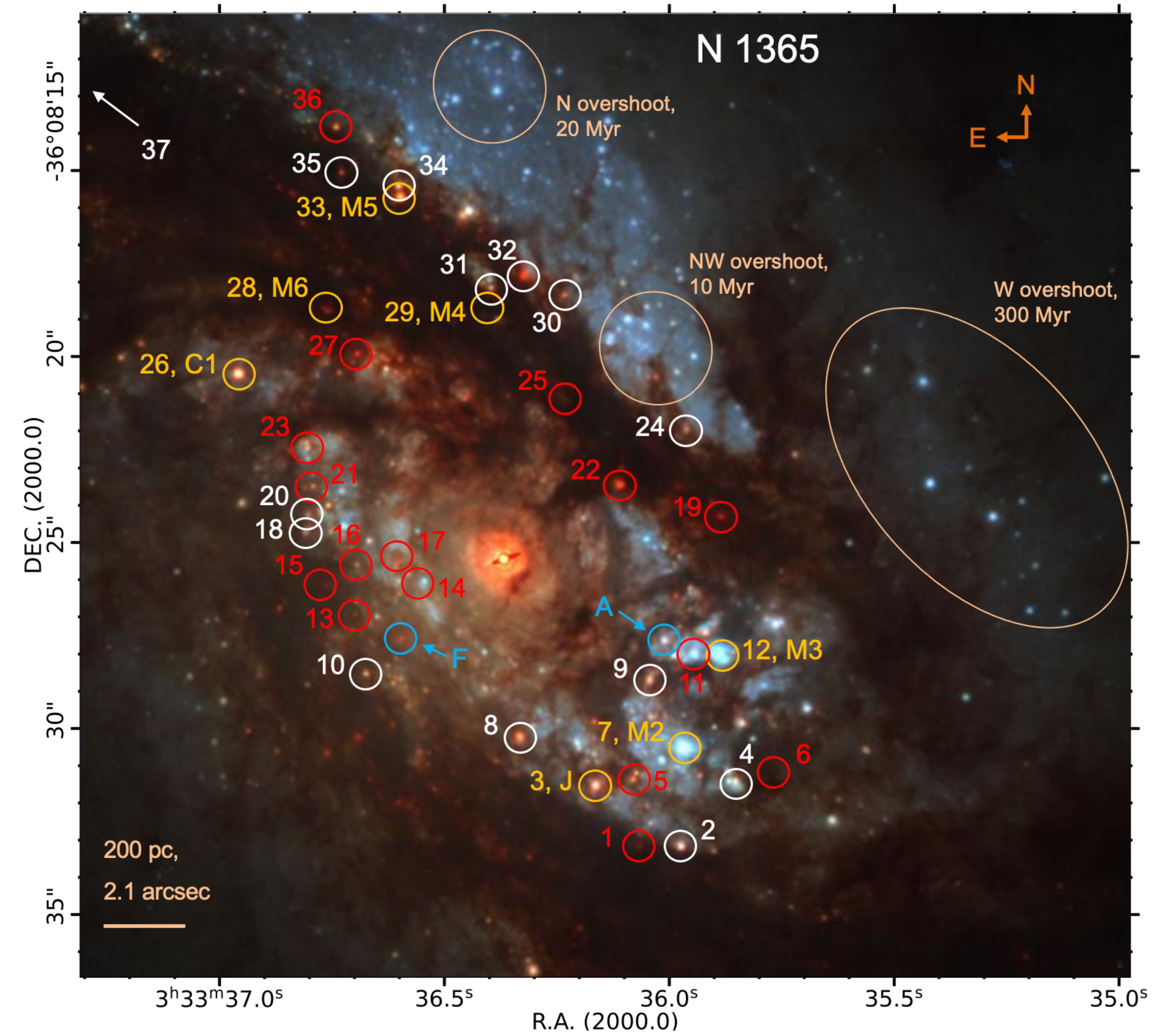}

\end{center}
\caption{IVB Hubble image of the central region of NGC 1365 showing the location of the massive young star clusters from Table \ref{tab:table_1}. Note that 36 of the 37 objects are in this central region, and only one (37) is slightly further up the dust lane along the bar.
White circles show the clusters from the Hubble compact clusters catalog. 
``Historical" clusters are identified by the gold circles. 
Red circles show the new clusters discovered in the JWST images.
The locations of radio continuum sources ``F" (believed to be related to the AGN jet) and ``A" (believed to be related to a supernova remnant  - see \citealt{sakamoto07}) are circled in blue, but are not included in the 37 clusters listed in Table \ref{tab:table_1}. Three overshoot regions discussed in Section \ref{sec:expand} are also identified. 
 }
\label{fig:reg_1_finding}
\end{figure*}

Figure \ref{fig:full_galaxy} shows the locations
of 
several parts of NGC~1365 that will be discussed in this paper. Our primary focus will be on the central 
$28\arcsec \times 26\arcsec$, or 2.7 kpc $\times$ 2.5 kpc, at the assumed distance of 19.57~Mpc, denoted as the ``central finding chart'' in Figure \ref{fig:full_galaxy} and shown in more detail in Figure \ref{fig:reg_1_finding}.
Region~1 is in the northern part of this region, 
where some of the most massive young clusters in the nearby universe have formed (\citealp{galliano08}, \citealp{galliano2012}).

\begin{figure*}
\begin{center}
\includegraphics[width =6.5in , angle= 0]{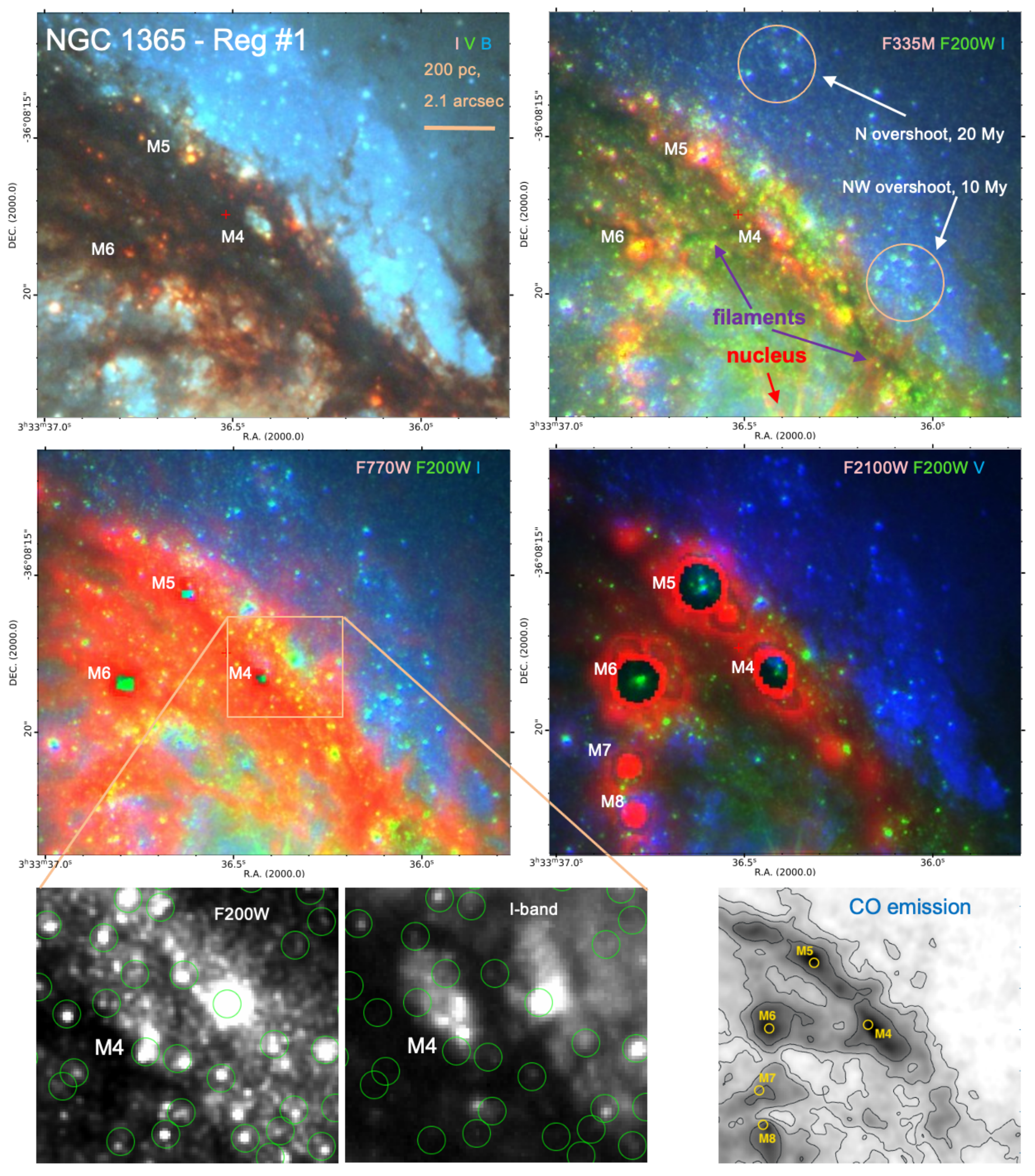}

\end{center}
\caption{
Region 1 in NGC 1365.
The figure shows an IVB image from Hubble in the upper left, and three images with different JWST bands  (F335M, F700W, or F2100) in red, F200W  (NIR continuum) in green, and I or V from Hubble in blue in the other panels. 
This region is just north of the nucleus, and contains the three strongest radio continuum sources in this galaxy, which appear to be young 
star clusters with masses of roughly log Mass in the range 6 to 7, 
(referred to as M4, M5, M6 by \citealt{sandqvist95}, \citealt{galliano08}, \citealt{galliano2012}).
Several thin dark filaments are shown along a ridge line containing some of the most recent star cluster formation in the F335M image. The holes in M4, M5, and M6 in both the F2100W and F770W images show where the data is saturated.
The largely unobscured region to the upper right shows  an ``overshoot region" outside the inner part of the galaxy
(see \citealt{sormani20}), with most of the clusters having estimated ages in the range between 10 to 20  Myr. The two bottom left panels show the excellent resolution of the JWST F200W image, and its ability to cut through the dust better than the I band Hubble image, which is shown to the right.  The bottom right panel shows the CO (2-1) line intensity emission map (Schinnerer et al. 2022, this volume) for Region 1. The small red cross in the top four panels is an artifact of the Hubble Legacy Archive which we used to make the figures, and should be ignored.   
Linear diagonal artifacts from the saturated nucleus are visible at the bottom edge of the three color panels using JWST filters.
}
\label{fig:reg_1}
\end{figure*}


\subsection{A Compilation of Massive Young Clusters in NGC 1365}
\label{sec:compilation}

In order to build a complete census of massive, young clusters in NGC~1365, we start by identifying all optically selected clusters that have estimated ages younger than 10~Myr and masses higher than $10^6~M_{\odot}$, from the cluster catalog and age-dating analysis based on multi-band HST optical images presented in \citealt{turner21} and  Whitmore et al. (2022, submitted). Three clusters are removed from the original list since they have very faint F2100W magnitudes indicative of old populations, unlike the rest of the young clusters in our final sample that are all very bright in F2100W.
We find 19 clusters which satisfy our criteria (these are the 19 clusters in Table \ref{tab:table_1} that have Cluster ID's in column 2).

To this list, we add 2 ``historical" clusters (M4 and M6) that were previously discovered in radio continuum images (e.g., \citealt{sandqvist82},
\citealt{sandqvist95}), with estimated ages less than 10 Myr and masses in the range 3~$\times~10^6$~to~$10^7~M_{\odot}$ (\citealp{galliano08}, \citealp{galliano2012}). 
 These are comparable to the most massive, recently formed clusters in the merging Antennae \citep{whitmore10} and NGC 3256 (\citealp{zepf99}, \citealp{mulia16}, \citealp{adamo21}) galaxies.

 Finally, we add 16 new, partly or heavily embedded massive young clusters identified in the new JWST images. These  have been selected to be brighter than 20.0 mag in at least one of the three filters which contain PAH emission (F335M, F770W and F1130W). As shown in Table \ref{tab:table_1}, many sources meet this criterion in all three filters, but a few only qualify in one or two. 
 
 We determined the magnitude limit of 20.0 mag to be included in our sample  by requiring that 
 the 19 clusters detected from HST (with one or two exceptions) be included. As will be discussed in Section \ref{sec:size}, 
 we also require the F200W magnitude to be brighter than 20.0 mag, to insure there is a bright stellar continuum source present, rather than just a dense cloud of gas/dust energized by nearby sources.
 We also require that the Concentration Index (CI - defined as the magnitude difference measured in 1 and 4 pixel aperture radii)
  in the F200W band  to be larger than that of a point source: CI$_{\rm F200W} > 1.4$~mag.

 Different cluster selection criteria are possible, and in fact several other works within the PHANGS consortium have selected clusters using a single filter: the F335M filter 
 \citep{rodriguez22}, F1000W (Schinnerer et al. 2022, this volume), and the F2100W filter \citep{hassani22}.
 In a future paper we will examine the impact of using different selection strategies in more detail.
 
 Our final census of 37 massive (Mass~$\gea 10^6~M_{\odot}$), young (Age $\lea$ 10~Myr) candidate clusters in NGC~1365 is compiled in Table \ref{tab:table_1} and shown in Figure \ref{fig:reg_1_finding}. Most of the clusters are in an inner star forming ring, as discussed by Schinnerer et al. (2022, this volume), while a few more are in the bar just to the NE of this ring (i.e., objects 33 through 37).

We find that NGC~1365, at a distance of about 20 Mpc, has the richest population of massive young clusters of any known galaxy within 30 Mpc, with about 30 clusters in this mass and age range compared with roughly a dozen in the merging Antennae galaxies (\citealp{whitmore10}), located at $\sim22$~Mpc.
  More distant galaxies with comparable populations are NGC~3256 
  (65 clusters, distance of 38 Mpc) and NGC 3690 (27 clusters, distance of 42 Mpc), according to \citep{Adamo20}. Both of these are ULIRGs and on-going merging systems, with estimated SFRs $\approx40$ M$_{\odot}$~yr$^{-1}$, roughly twice as large as  in NGC~1365 \citep{lee22}.

We are now ready to address the question: ``how many massive young clusters are so enshrouded in dust that they are missing at optical wavelengths?" From column 11 of Table \ref{tab:table_1} we find only 4 out of 37 (11\%) of our cluster catalog are so dusty that they are not clearly visible in the I-band image (longest wavelength observed from HST).  
However, age and mass estimates are quite challenging without photometry in at least three optical bands, which means clusters that are not also detected in the B and V filters are unlikely to be included in optically-selected cluster catalogs, and hence might be considered missing in some cases.  For NGC~1365, we find that 9 of 37 ($\sim24$ \%) clusters are ``missing" in the V band, and 11 of 37 ($\sim30$ \%) are ``missing" in the B band.  {\em We therefore find that between $\sim11-30$\% of the clusters are ``missing" from optical catalogs because they are embedded at some level.}  This is similar to the situation in the Antennae, where only 2/13 = 15 \% of bright thermal radio continuum sources were ``missing" at optical wavelengths, and 38 \% of fainter clusters (with S $>$ 70 $\mu$Jy)  were missing \citep{whitmore02}.

\begin{figure*}
\begin{center}
\includegraphics[width =7in , angle= 0]{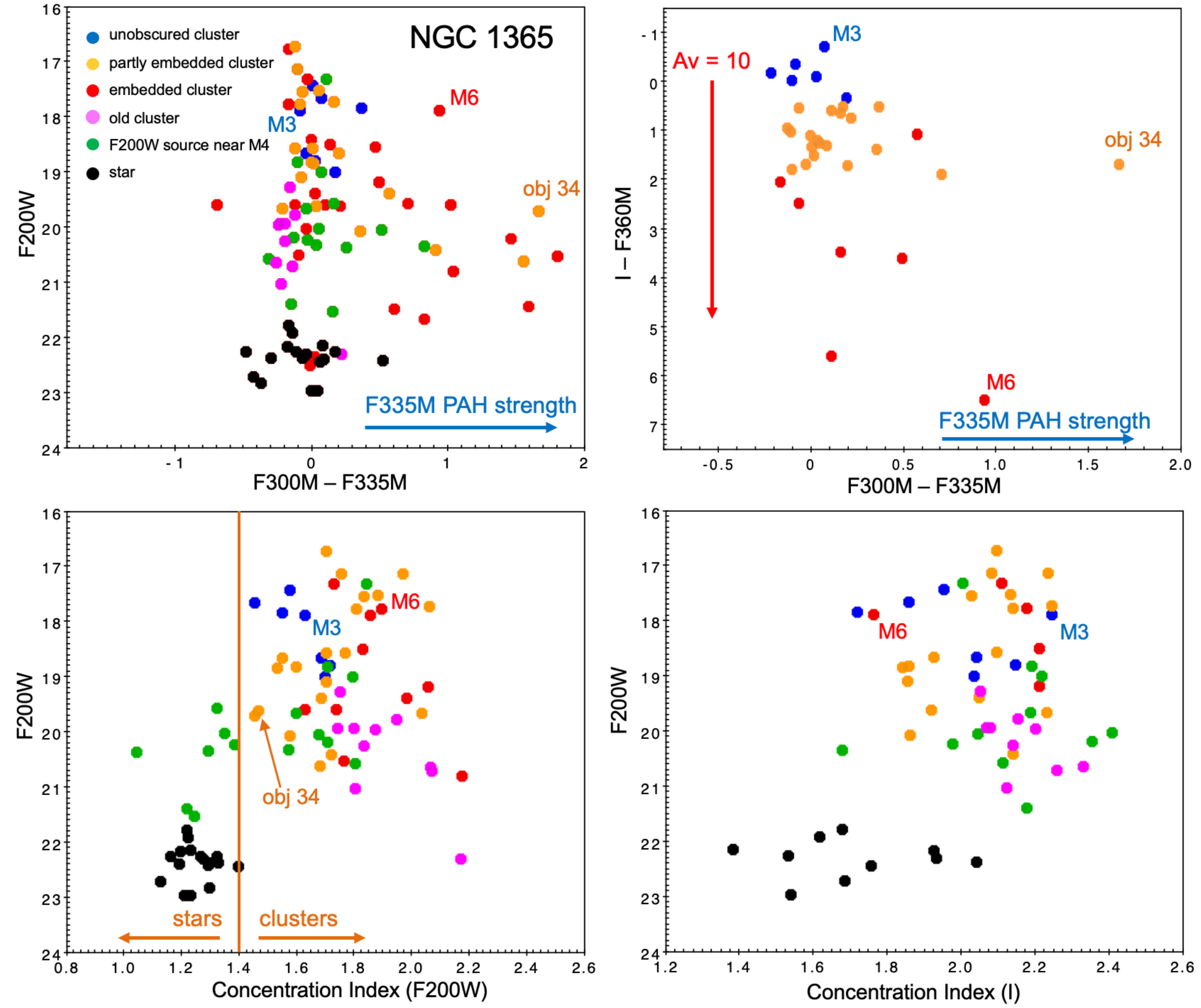}

\
\end{center}
\caption{Diagnostic diagrams for clusters and stars in NGC 1365.
Top Left: - F300M - F335M vs. F200W color-magnitude diagram including unobscured clusters (blue circles), partly embedded clusters (orange), embedded clusters (red), old clusters (pink), F200W sources near M4 (both clusters and stars; green), and stars (black) from the outer part of the galaxy. Top Right:
 F300M - F335M vs. I - F360M color-color diagram, using the same color coding.
Bottom Left - Concentration Index measured from the F200W filter.  The vertical  orange line shows that stars and clusters nicely separate at 
CI$_{F200W} = 1.4$.
Bottom Right: Same diagram as to the left, but now using CI measurements in the I filter, which do not distinguish stars from clusters as well as the JWST F200W filter.
 }
\label{fig:size}
\end{figure*}

Figure \ref{fig:reg_1} reveals a number of interesting features in Region~1. The top-left panel shows the strong dust lane cutting through an optical IVB image taken with HST. Three different JWST$+$HST filter combinations (as indicated along the top right corner of each panel) reveal embedded young clusters, some of which are very bright and saturated (e.g., M4, M5, M6 in the F2100W and F770W images), as well as some blue, slightly  older clusters outside the dust lane.
All three of these saturated clusters are also strong radio continuum sources (corresponding to sources D, E, G, respectively in \citealt{sandqvist95}). 

There are also very thin, dark filaments visible in the F335M/F200W/I-band image; a few are labeled in Figure \ref{fig:reg_1}.
These are typically found along the inside edge of some of the strongest F335M PAH (Polycyclic Aromatic Hydrocarbons) ridges, and have widths of a few tens of parsecs and lengths of a few hundred parsecs, similar to the massive filaments found in the Milky Way \citep{hacar2022}, normal spiral galaxies (e.g., Thilker et al. 2022, this volume, \citealt{Meidt22}) as well as other star bursting galaxies such as the Antennae \citep{whitmore14}.

\subsection{Estimate of Star Cluster Properties}
\label{sec:properties}


Figure \ref{fig:size} presents four diagnostic diagrams that will be used in this section to estimate several basic properties of the clusters. These include a color-magnitude diagram, color-color diagram, and two flux vs. concentration index diagrams using different bands. 
A visual estimate of the level of embeddedness is included in Table \ref{tab:table_1}: Class 1 = largely unobscured (A$_V \lea 3$), Class 2 = partly embedded  ( 3 $\lea$ A$_V \lea$ 10), and   Class 3 = embedded (A$_V \gea 10$). 
The location of data points for three clusters are shown in each panel to help demonstrate the properties of a largely unobscured cluster (M3, shown in blue), a partially embedded cluster (obj 34, shown in orange), and an embedded cluster (M6, shown in red). Arrows  show how properties track in the diagram, including PAH emission strength (youngest clusters), extinction (dustiest clusters), and size (to help separate stars from barely resolved clusters). 

We create several different training sets to illustrate a number of features.
These include the 37 massive young clusters discussed in Section \ref{sec:compilation}, now broken into three subsamples:
unobscured, partly embedded and embedded. Other training sets include older clusters (with ages around 300 Myr), a set of largely embedded F200W sources in the dust lane around M4 (see Figure \ref{fig:reg_1}), and objects that are likely to be red super giants (RSG) in a stellar association near the eastern end of the bar. 
The color coding for the different samples is shown in the upper left panel of Figure \ref{fig:size}.

In the next subsections, we show how we separate individual stars from clusters, and then describe methods to estimate the extinction (Section 3.2.2), mass (Section~3.2.3), and age of embedded clusters (Section~3.2.4).  We adopt the estimated 
properties of relatively unobscured clusters determined in Whitmore et al. (2022, submitted) using Hubble data, as included in Table \ref{tab:table_1}, to establish correlations that allow us to estimate properties for the more deeply embedded clusters observed primarily in the JWST bands.   

\subsubsection{Separation of Clusters and Stars}
\label{sec:size}

The bottom two panels in Figure \ref{fig:size} show that the F200W filter provides a powerful new way to differentiate stars from barely resolved clusters, and to eventually estimate the sizes of the clusters. 
The scatter in the Concentration Index for stars (black points) using the F200W filter on JWST is dramatically less than from the I filter on HST. If we use the V filter from HST instead the  results are intermediate between F200W and I, but we are not able to penetrate the dust as well. This better resolving power might be expected since the resolution with JWST in the F200W band is approximately~$0.05\arcsec$, while it is approximately~$0.08\arcsec$ in I using Hubble. In addition, the sampling is better with $0.031\arcsec~\mbox{pix}^{-1}$ for the F200W filter  compare to $0.040\arcsec~\mbox{pix}^{-1}$ for the I filter, and the former provides these measurements even in dusty regions. 

Figure \ref{fig:size} shows that while there is a large overlap between stars and clusters in CI$_{\rm I}$, there is no overlap using the CI$_{\rm F200W}$ measurements, and the orange line is able to cleanly separate stars from clusters in all cases for our training sets. In the future, we plan to use this new capability to improve the size measurements of star clusters as well. 

This improved performance also shows that while some of the embedded F200W objects (green points) in the bottom left panel of Figure \ref{fig:reg_1} are likely to be stars, most (particularly the brighter ones) appear to be clusters, probably with 
Mass in the range 10$^5$ to 10$^6$ M$_\odot$.

\subsubsection{Extinction Estimates for Embedded Clusters}
\label{sec:extinction}

There are two approaches we can take to estimate extinction. The first is to use the Galactic extinction law from \citep{Fitzpatrick99}. 
The I - F360M vs. F300M - F335M diagram  shown in the upper right panel of Figure  \ref{fig:size} 
demonstrates the range of extinction (A$_V$) values found in our sample.
A reddening vector following a Galactic extinction law  is shown for $A_v = 10$~mag.
We find that this predicts 4.75 mag of reddening in I - F360M, hence the coefficient connecting $A_v$ and I - F360M is 10 / 4.75 = 2.11.
Very small deviations from exactly vertical are ignored since the horizontal dimension is primarily an indication of PAH emission strength rather than reddening. For example, both M6 and obj 34 are found in regions with strong F335M emission, explaining their positions on the right side of both the color-magnitude and color-color diagrams.
We find that the majority of the clusters have A$_V$ values less than 10~mag.
The underlying assumption needed to estimate extinction using the I - F360M color index is that all the (young) clusters in our sample have similar intrinsic colors, hence the redder color is caused by extinction rather than due to age. The fact that unobscured clusters (blue points) all have I- F360M values around 0 supports this assumption.

A second approach is to use the empirical correlation between the E(B-V) values from the SED-fitting based on Hubble data vs. I - F360M, as shown in the bottom panel of Figure \ref{fig:mass_ebv_2plot}. 
We use the sample of class 1 and 2 (symmetric and asymmetric clusters, human classified sample - see \citealp{whitmore21}) in NGC 1365 to make this determination. 
\citet{turner21} describe the basic SED age-dating method used in the PHANGS-HST pipeline. More recently, Whitmore et al. (2022, submitted) describe modifications which identify and correct the age estimates for old globular clusters in the pipeline; many old globular clusters have incorrect age estimates in the default pipeline due to challenges in disentangling age/reddening/metallicity. 
This helps remove interlopers that might otherwise pollute our young clusters sample. The fraction of clusters with incorrect age estimates in NGC~1365 are reduced from $\approx20$ \% to $\approx10$ \% according to  Whitmore et al. (2022, submitted).

Only the log Age $<$ 6.7 clusters are included to make the sample similar to our program sample of young massive clusters in NGC 1365. The RMS scatter (0.95 mag) is relatively large and the correlation coefficient (0.49) is only fair, hence our estimates of extinction will only be approximate. 
The resulting formula is: 
$A_V (\mbox{JWST) = [(m$_I$ - m$_{F360M}$)}  \times 0.810 + 0.99] \times 3.1$, which provides estimates of extinction for our partially embedded clusters. 
Hence the coefficient connecting $A_v$ and  I - F360M is 0.810 $\times$ 3.1 = 2.51, relatively similar to the estimate from the Galactic extinction law discussed above (i.e., 2.11). This is the formula we will use in Table \ref{tab:table_1}.

In principle it would be advantageous to use only JWST observations to determine the correlation, since four of our clusters do not have I band photometry.  However, we found that the smaller baseline resulted in considerably more scatter, and hence was not practical.
We note that the relationship we derive is specific to the analysis presented here since aperture corrections have not been made to our data yet; other researchers will need to determine the constants from their own photometry.

Table \ref{tab:table_1} shows that the highest estimated extinction value is for the intrinsically bright object 28 (= M6/G, e.g., saturated in the F770W and F2100W filter observations, as shown in Figure \ref{fig:reg_1}), which has an estimate of A$_V$ = 19.3 mag. 
The value would be about A$_V$ = 15 mag if the Galactic extinction formula was used instead.
This extinction is even larger than  WS80, a deeply embedded star cluster in the Antennae galaxies \citep{whitmore02} that is the strongest CO and 15 micron source in that system, and has an estimated age of 2 Myr, A$_V\approx8$~mag, and  log Mass$~\approx 6.6$. This indicates that the very different environments of a chaotic merger, and the central region of a barred spiral galaxy, can both produce very massive, heavily extincted, young clusters.

\subsubsection{Mass Estimates for Embedded Clusters}
\label{sec:mass}


To estimate masses for our embedded clusters we use the approach developed by \citet{rodriguez22} who show that there is a good correlation between the F200W flux and the HST-based mass estimates for clusters in NGC~7496, when limited to very young clusters. A similarly good correlation is seen for F335M. 
This is because similar age clusters have similar mass-to-light ratios. 

\begin{figure}
\begin{center}
\includegraphics[width =3.3in , angle= 0]{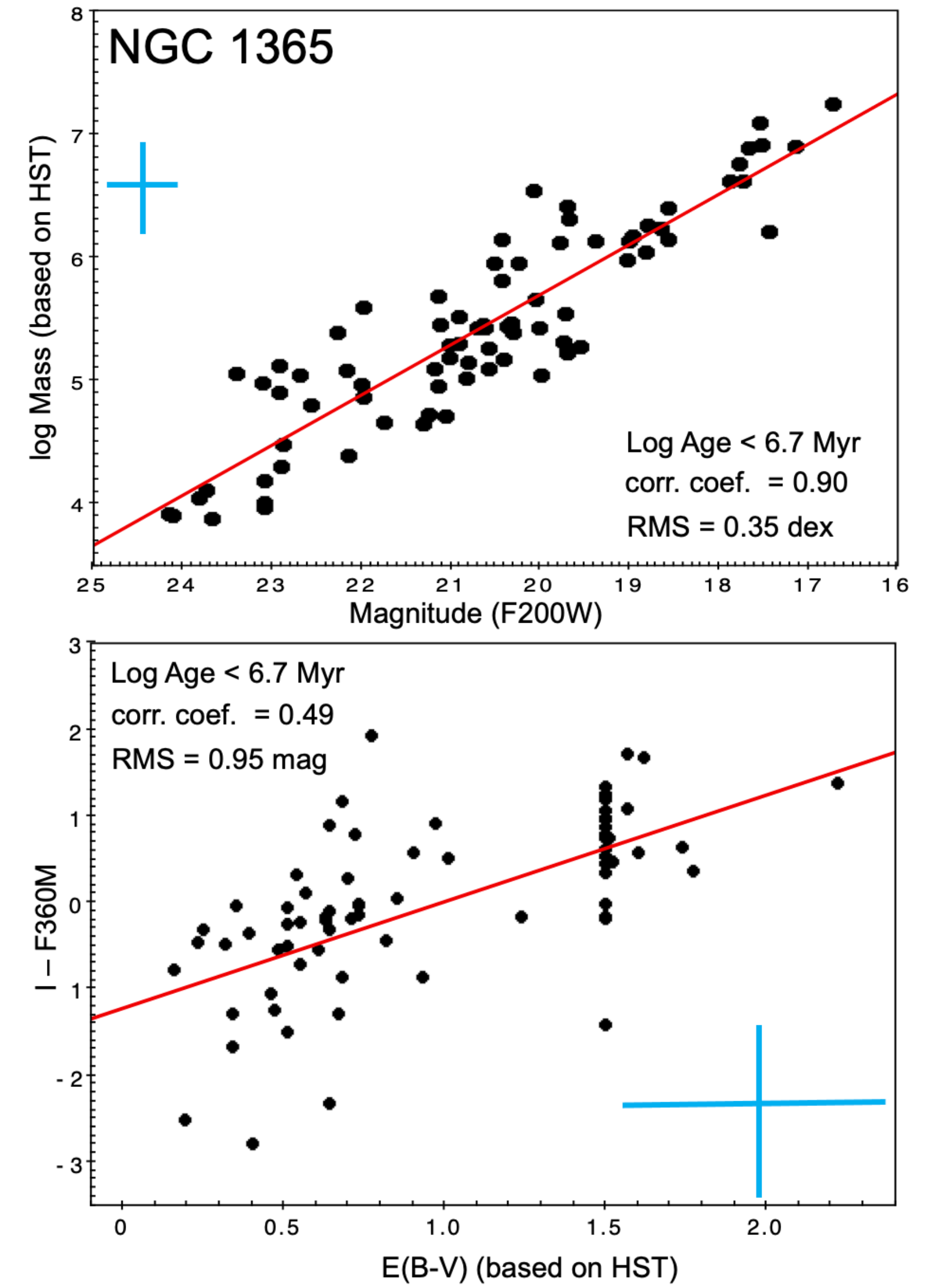}
\
\end{center}
\caption{The top panel shows the correlation between the JWST F200W magnitudes and the HST-based cluster mass estimates from Whitmore et al. 2022 (submitted) for class 1 (symmetric) and 2 (asymmetric) clusters for a sample with log Age $<$ 6.7. The bottom panel shows the relationship between I - F360M magnitudes and the HST-based reddening (E(B-V)) estimates for the same sample. The correlation coefficient and the RMS scatter are provided, and the crosses show the typical uncertainty.   }
\label{fig:mass_ebv_2plot}
\end{figure}

As shown in the top panel in Figure \ref{fig:mass_ebv_2plot} we  find a very good correlation in NGC1365 as well, with a correlation coefficient of 0.90, and an RMS scatter of just 0.35 dex. The resulting formula is  
$\mbox{log}~(M/M_{\odot}) = - 0.4068 \times m_{\rm F200W} + 13.83$. We again use the sample with  log Age $<$ 6.7 clusters to make it similar to our sample of young massive clusters in NGC 1365. 
A version of this fit based on the F300M magnitude is nearly as good: $\mbox{log}~(M/M_{\odot}) = - 0.3639 \times m_{\rm F300M} + 13.25$, with correlation coefficient = 0.87. 
If the mass-to-light ratios were exactly equal for all the objects we would find the coefficient in the formulae would be the inverse of 2.5 in the definition of magnitudes (i.e., -2.5 log flux), hence it would be 1~/~2.5 = 0.40. We note that this is true, with our empirically determined coefficients approximately equal to 0.41 and 0.36.
The final mass values included in Table  \ref{tab:table_1} use the mean of these two mass estimates.



Our estimates for clusters M4, M5, and M6 range from log Mass = 6.0 to 6.6, in fair agreement with \citet{galliano08} who estimate values ``on the order of 10$^7$~M$_{\odot}$" for the three clusters. We note that in \citet{galliano2012}, the estimates are slightly lower, ranging from log Mass = 6.5 to 7.0,
which are in better agreement with our values.  

 We find that 
 7 of the 37 clusters in Table~1 have estimated masses that are somewhat below log Mass = 6.0. This is because there are uncertainties at the factor of $\sim2-3$ level associated with each cluster mass.  Going forward, for various estimates we assume that 30 clusters from our sample have masses of at least log Mass $\gea$ 6.0.

 \subsubsection{Age Estimates for Embedded Clusters}
\label{sec:ages_for_embedded}

In the future, we will combine HST and JWST bands to improve age estimates, especially for partially embedded clusters.
For now, we will use the same basic approach as in the previous two subsections
to estimate the ages of the embedded and partly embedded clusters based on HST estimates. 

\begin{figure}
\begin{center}
\includegraphics[width =3.3in , angle= 0]{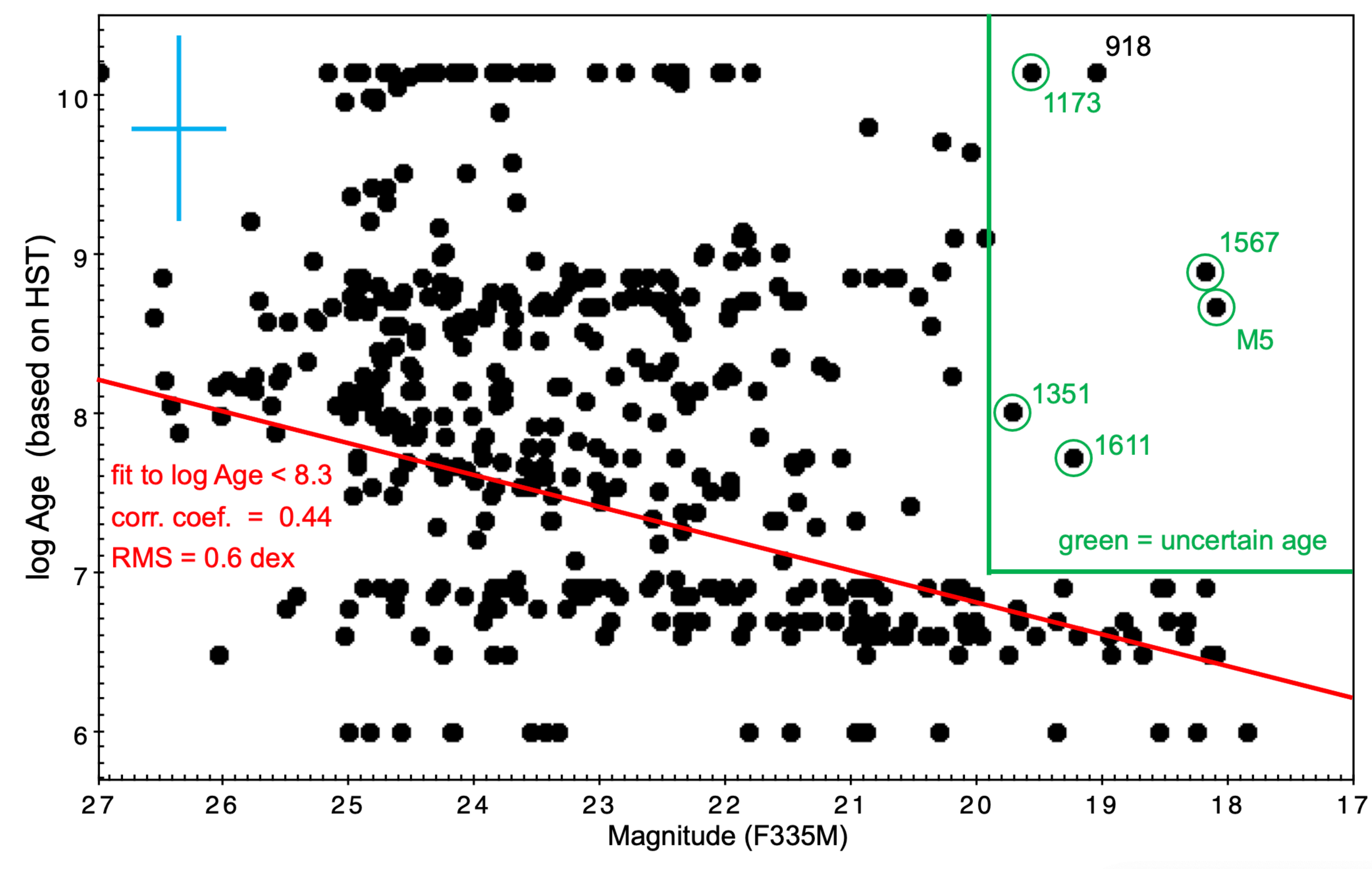}
\
\end{center}
\caption{A plot of JWST F335M magnitude versus age estimates based on Hubble data from Whitmore et al. 2022 (submitted) for class 1 (symmetric) and class 2 (asymmetric) clusters.  This shows a large degree of scatter and essentially no correlation for clusters with log Age $\ge$ 8.3. However, we find that most of the clusters with F335M brighter than 20 mag have log Age values less than $\approx7$ (10 Myr). A visual examination of the objects in the upper right rectangle shows that all but one of the clusters (i.e., the objects with green circles around them) have H$\alpha$ emission nearby, indicating that the older age estimate is probably incorrect. The blue cross shows the approximate uncertainty based on a fit to the points with log Age $\le$ 8.3, shown as the red line.} 
\label{fig:age_335}
\end{figure}

As shown in Figure \ref{fig:age_335}, the age estimates are somewhat more complicated than extinction and mass estimates,  with more scatter and a nonlinear dependence.  This figure plots HST-based ages vs. source brightness in the F335M filter, which includes the 3.3$\mu$m PAH feature and also continuum emission.  While there appears to be a possible correlation for ages between log Age = 6 and 8.3 (e.g., correlation coefficient = 0.44), there is essentially no correlation at older ages.

However, at the bright end (i.e., with F335M $<$ 20.0 mag) we find that most sources are young, with log Age less than 7. A  visual examination of the clusters in the upper right part of the diagram show that five of the six (circled) are in regions with strong H$\alpha$ emission and are likely to be young; hence the HST-based age estimates are likely to be incorrect. An example is M5 (object 28), which is a very strong radio source which has been age dated using spectral observations indicating that it is clearly younger than 10 Myr \citep{galliano2012}.
We note that the plots for other PAH bands, notably F770W and F1130W, look very similar to that for F335M. 

Based on this evidence, we  believe at least 90~\% of the objects with F335M brighter than 20 mag are younger than 10 Myr. We note that this would be compatible with the results from Whitmore et al. 2022 - in prep) who find that approximately 10~\% of the HST-based age estimates of clusters in the inner region of NGC 1365 are overestimated, due to the large amount of dust that makes the clusters appear red, and potentially old. While low-resolution (0.8 arcsec) H$\alpha$ imaging from MUSE \citep{emsellem22}  has been used in Whitmore et al. 2022 (submitted) to improve the age dating to some extent, this is not as good as having high resolution Hubble H$\alpha$ imaging. 

\begin{figure}
\begin{center}
\includegraphics[width =3.3in , angle= 0]{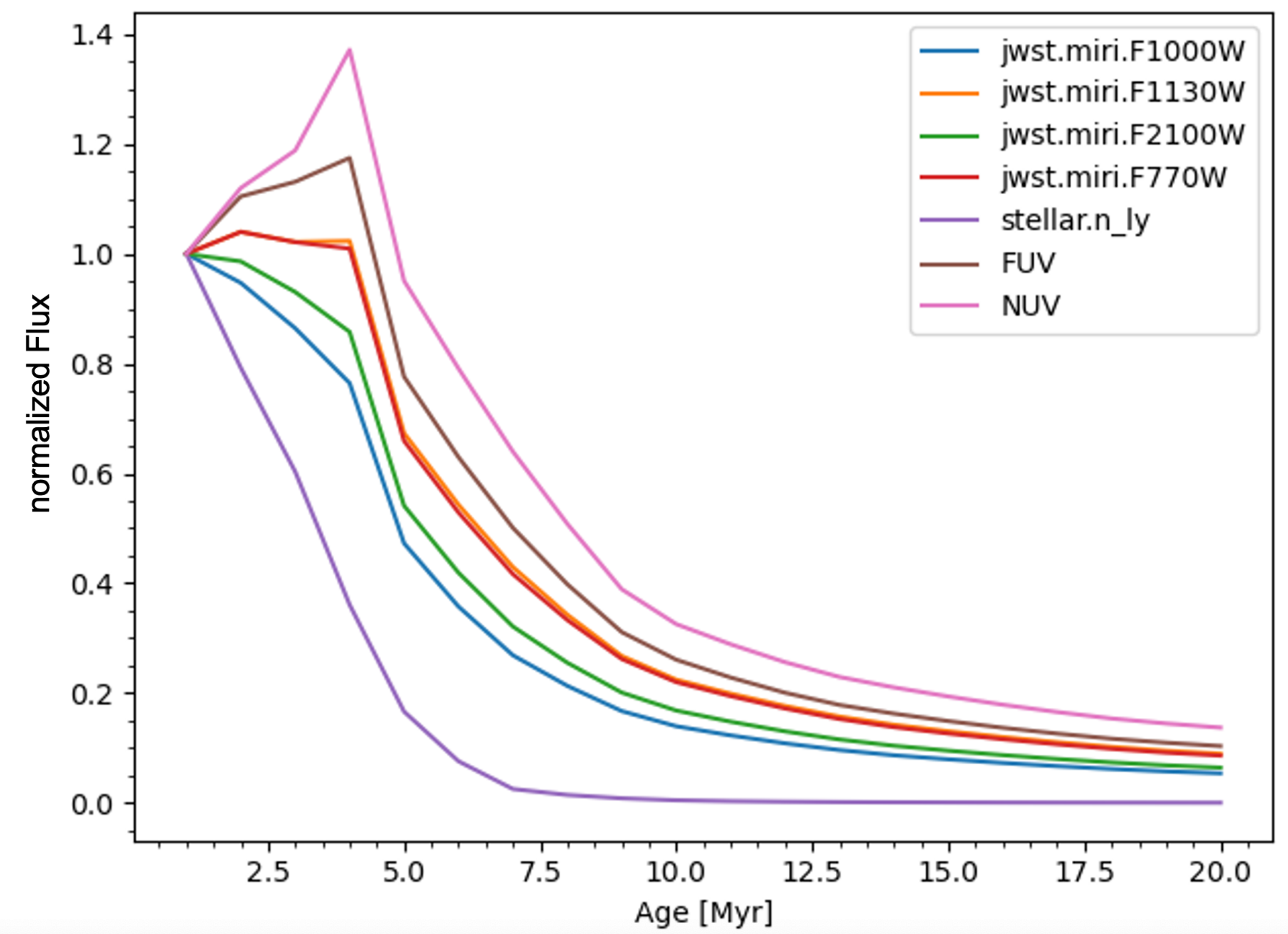}
\
\end{center}
\caption{The predicted evolution of the flux emitted from clusters in a variety of filters, from CIGALE \citep{Boquien19}. We include the JWST MIRI F770W, F1000W, F1130W, and F2100W filers, each including contributions from the nebular emission, which is an important component (i.e., without the nebular emission the various MIRI bands are nearly on top of each other); the   NUV  and FUV filters from GALAX; and the nebular emission alone (i.e., proportional to Ly$\alpha$ in this case but similar to other line emission such as H$\alpha$). A legend is provided in the upper right. The profiles are normalized to 1.0 at an age of 1 Myr. The key point is that strong emission in the JWST bands is generally indicative of young ages,  but the drop off is not as steep a function of age as for the nebular lines.  }
\label{fig:pah_profiles}
\end{figure}

Figure \ref{fig:pah_profiles} shows there are physical reasons why we expect most of the brightest PAH emitters to be young. This figure shows a normalized  (at 1 Myr) plot of predicted flux values in various MIRI bands from the Draine models of interstellar dust emission (\citealp{draine21}). There are similar profiles for the NUV and FUV emission that are the main driver of the PAH emission. In all these cases the peak flux occurs for ages around 2 to 4 Myr, with relatively rapid declines within just 10 Myr.  

As expected, the nebular emission (purple line) shows an even more rapid drop off, falling to just a few percent by 6 or 7 Myr. This is why nebular lines, such as H$\alpha$, are so valuable for age dating; if you see strong emission associated with a cluster, the source is young (e.g., \citealp{whitmore02},  \citealp{anders03}, \citealp{chandar10a}, \citealp{fouesneau12}). The PAH band strengths have a similar predicted time evolution, although they typically drop to only  $\sim20$~\% of their initial flux by 10 Myr. This is why the PAH emission is  
correlated with young cluster ages in Figure \ref{fig:age_335}. We note that this drop off in PAH strength as a function of time also appears to be  compatible with the results of \cite{lin20} for LEGUS galaxies. 

Since the scatter in the relationship between F335M magnitudes and log Age based on HST is fairly large ($\sim$0.6 dex), we have chosen to simply assign log Age values of 6.5 (3 Myr) for the clusters without HST age estimates. This is consistent with the results from Figure \ref{fig:age_335}, the flux profiles in Figure \ref{fig:pah_profiles}, and the comparison with Hubble age estimates in Table \ref{tab:table_1}. Hence, our final age estimates are included as the second entry in column 6 in Table \ref{tab:table_1}, and are either the HST-based SED fitting value if it exists, or log Age = 6.5 (3 Myr) for the clusters without estimates from Hubble.

\subsection{Duration of the Embedded Phase}

A complete census of massive clusters from the optical and infrared allows us to estimate the time that clusters remain in the deeply embedded phase, specifically the length of time they are not detectable at optical wavelengths.  While early estimates (e.g., \citealt{kawamura09}) were longer than 10 Myr, more recent values have generally been in the range 1 - 5 Myr 
(e.g.,  \citealt{lada03}, \citealt{whitmore14}, \citealp{hollyhead15},
\citealt{grasha18}, \citealt{chevance20}, 
\citealt{kim21}, \citealt{hannon22}, \citealt{He22}, and Kim et al. 2022, this volume).
NGC~1365 provides a particularly interesting case because of the large amount of gas and dust which might, in principle, result in longer timescales.  It also provides a unique view of the embedded phase of extremely massive ($\geq 10^6~M_{\odot}$) clusters, which can eventually be compared with results from lower-mass clusters to assess whether or not the duration of this phase has any dependence on cluster mass. 

We calculate the duration of the deeply embedded phase for massive clusters in NGC~1365 as:
\#missing $/$ total $\times$ Age (i.e., sample interval) = 4/30 $\times$10~Myr = 1.3 $\pm$ 0.7~Myr, a very short timescale.  
If we instead consider the timescale for massive clusters to be detectable in the B, V, and I bands (a minimum needed for most optically-based age-dating methods), we find a somewhat longer timescale 
\#missing $/$ total $\times$ Age = 11/30 $\times$~10~Myr = 3.7 $\pm$ 1.1~Myr.  

 Our estimates for how long very massive clusters in NGC~1365 remain embedded in their natal ISM are consistent with those found by previous works, without being obviously longer or shorter relative to estimates made for lower mass clusters in other galaxies.   It is important to establish if the length of the embedded phase depends on cluster mass or not, because this provides direct constraints on early feedback processes which may be important for 
limiting the masses of clusters.

\begin{figure}
\begin{center}
\includegraphics[width =3.3in , angle= 0]{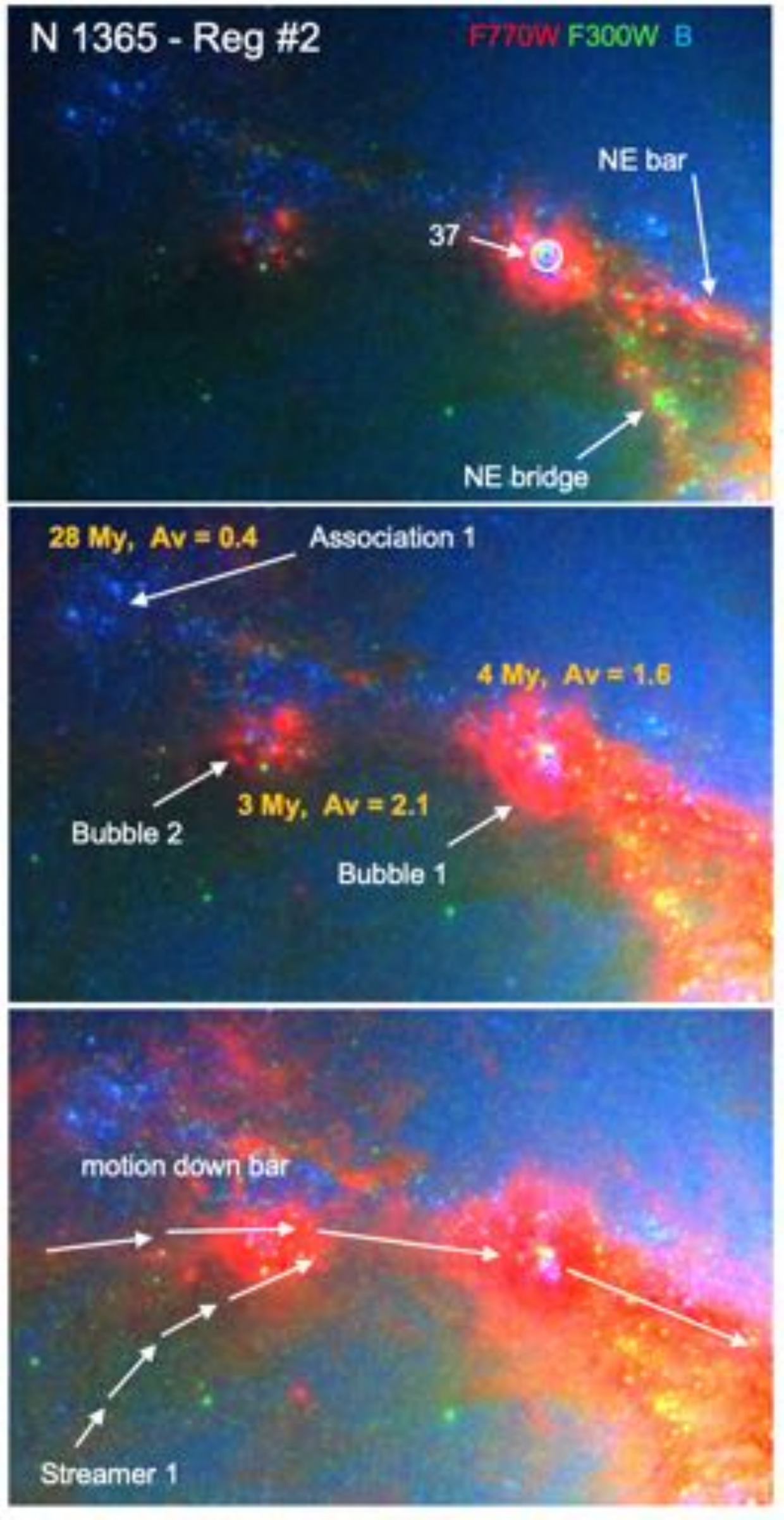}
\end{center}
\caption{ JWST (F770W and F300M) and Hubble (B band) images of Region \# 2, just up the NE bar from the nucleus, as shown in Figure \ref{fig:full_galaxy}. The different contrast levels highlight key features, including two young 'bubbles' resulting from star-formation that may have been triggered by a intra-bar streamer colliding with gas and dust in the bar itself, as shown by the arrows. Another feature, which is only visible in the JWST bands due to the heavy dust in the area, is the ``NE Bridge", as discussed in the text. } 
\label{fig:reg_2}
\end{figure}

\begin{figure*}
\begin{center}
\includegraphics[width =5.0in , angle= 0]{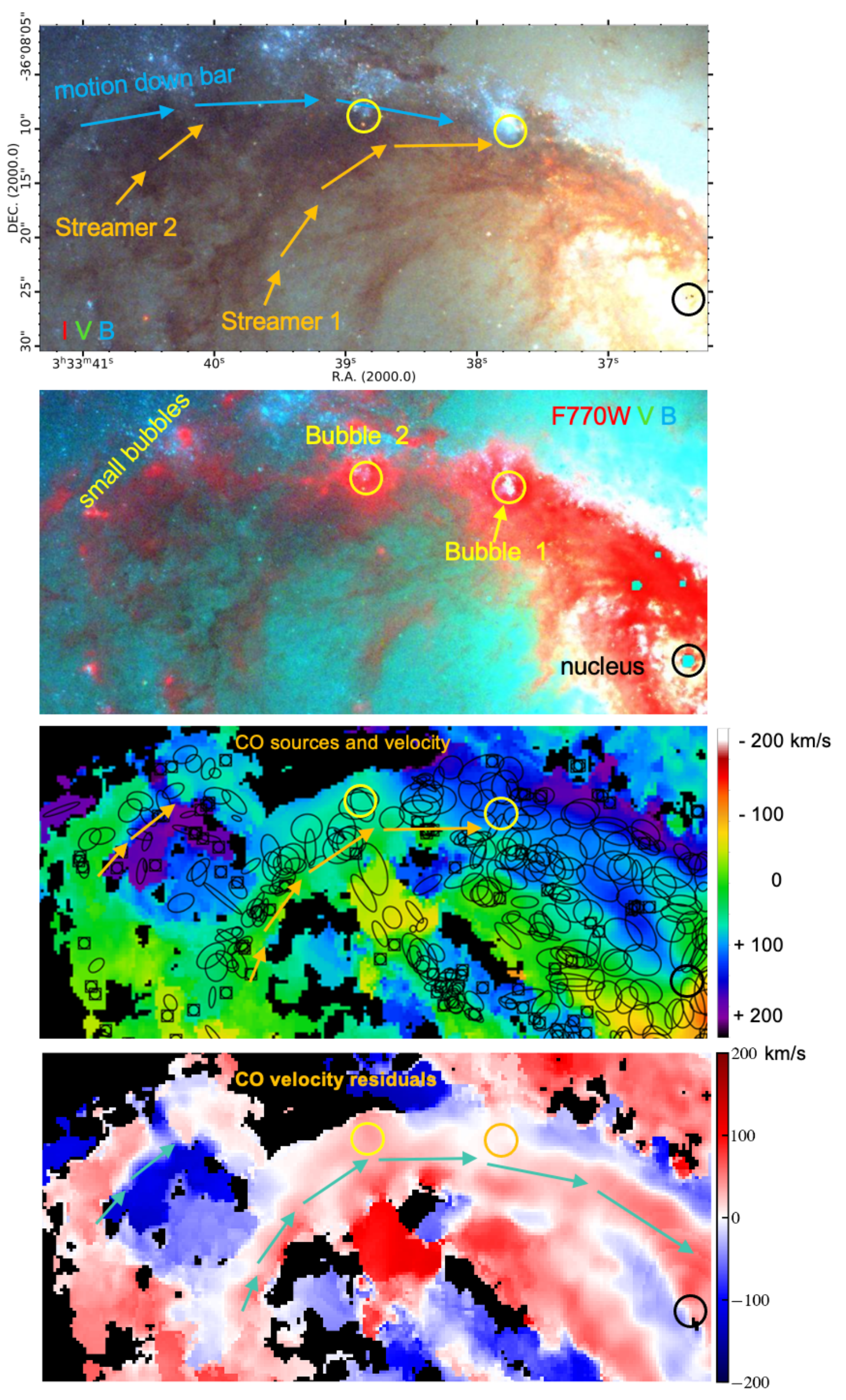}
\end{center}
\caption{Four images showing the extended Region 2 field of view. The top panel shows IVB observations from Hubble, and the second panel the combined JWST (F770W) and Hubble (V and B) images.  Objects identified in Figure \ref{fig:reg_2} and the location of Streamer 2 and the nucleus are included. The third panel
shows giant molecular clouds identified via their CO(2-1) emission (ellipses represent their fitted spatial extents), while the fourth panel shows velocity residuals from a model of the inner disk rotation 
from Schinnerer et al. 2022 (this volume). See text for discussion. } 
\label{fig:triggering_4plot}
\end{figure*}

\section{Region 2: Star Formation Triggered by Collisions of Gas and Dust Streamers}
\label{sec:reg_2}

\subsection{Triggered Star Formation}
\label{sec:triggered}

While the strongest star formation in NGC 1365 is occurring in the inner region (i.e., 36 of the 37 youngest most massive clusters as we saw in Section \ref{sec:compilation}),
a moderate amount is also taking place along the bar, an example of which is shown in Figure~\ref{fig:reg_2}.
In particular, Bubble 1 is object 37 from Table 1, with a log Mass  = 6.6. 
Much of the star formation in Region~2 appears to be triggered by dust and molecular gas from dark filamentary ``streamers" 
that reach the bar, colliding and merging with the gas and dust as illustrated in the lower panel of Figure \ref{fig:reg_2}, and as  discussed in \cite{elmegreen09}. 
Another excellent candidate for this type of triggered cluster formation 
is Bubble 2 (3 Myr, $A_V = 2.2$~mag; age and extinction estimates from Whitmore et al. 2022, submitted), as shown in Figure \ref{fig:reg_2}.
A careful look at Figure \ref{fig:full_galaxy} shows several other examples of regions of star formation near the intersection of streamers and the bar. 

 Previous kinematic work has shown strong streaming motions due to the bar in NGC~1365 (e.g., \citealp{sakamoto07},  \citealp{lindblad96},  Schinnerer et al. 2022, this volume).
Simulations
provide a more visual way to understand the dynamics of streamers of gas as they approach, interact with, and sometimes overshoot the bar lane. We encourage interested readers to watch this video.\footnote{https://www.youtube.com/watch?v=1elJ81GlIlA\&list=PLlsb6ZGKWbI6QOFPd6OYU4zLXCmBa6uxx\&index=5}  
Figure 11 in \citealt{sormani20} also provides insight to understand the dynamics and the role of streamers as they interact with the bar, potentially triggering star formation.
One key new result from JWST observations  is how ubiquitous these streamers are in spiral galaxies (see 
\citealt{Meidt22} and Thilker et al. 2022 this volume).

Figure \ref{fig:triggering_4plot} shows a larger field of view around Region 2 (i.e., ``Region 2 extended" from Figure \ref{fig:full_galaxy}) and contains additional kinematic information which can be used to help test the triggering hypothesis. A full kinematic analysis is beyond the scope of the current paper, however.  
The top panel shows a IVB Hubble image which highlights the dust lanes and regions of young stars. Besides the streamer we show in Figure~\ref{fig:reg_2}, which is labeled Streamer 1, we find several other candidate streamers. The second panel uses the F770W PAH emission filter (shown in the red) to highlight the youngest star forming regions. The third panel shows CO (2-1) moment-1 velocity from the PHANGS-ALMA observations \citep{leroy21}, color-coded from -220 km/sec (dark blue) to +220 km/sec (dark red and white). CO clouds are marked using ellipses and box-circle symbols.  Ellipses are sized to represent the deconvolved major/minor cloud axes (FWHM), whereas box-circles indicate unresolved clouds (Rosolowsky et al. 2022 - this volume).

We first note the large number of CO clouds in Streamer 1, indicating that much of the gas and dust feeding into the bar, and eventually into the inner region of the galaxy, is coming from the streamers rather than from the bar itself. 
This figure shows that the gas velocities along Streamer 1 are similar (i.e., this feature appears as a uniform, green color, indicative of velocities around 0 km/s), as expected for a coherent physical component.

 Gas velocities along the bar (the upper blue arrows) are more mixed, with values $\approx -100$ to $-200~{\rm km~s}^{-1}$ along most of the bar, but with velocities $\approx -50~{\rm km~s}^{-1}$  in regions where the streamers intersect the bar. This appears to be what is happening in the region of Bubble 2, with the velocities shown by a light blue-green color indicating a velocity that is intermediate between that of the bar and Streamer 1,  as we would expect if this formed from the collision of the two components. The situation is less clear for Bubble  1,  but the color appears to be light blue ($\approx -100~{\rm km~s}^{-1}$), as expected if  it is a combination of dark blue (just to the north and south) and green from Streamer~1.
 
 We also find a less CO-populated streamer farther out in the bar, which we have labeled Streamer 2. It appears to be associated with three or four small star formation regions rather than a large event like the formation of Bubbles 1 and 2. However, we find that it has a similar morphology with uniform green colors until it mixes with the bar, and then shows a combination of green and blue velocities beyond that point. 
 
 The bottom panel of Figure \ref{fig:triggering_4plot} shows the CO velocity residuals based on a model of the inner rotating disk (Schinnerer et al. 2022 - this volume). This supports the same basic interpretation, but provides an even stronger connection between Streamer 1 and the feeding of the inner region, with a relatively uniform, narrow red component  all the way into the central region.   We also note that the association of Bubble 1 with the streamer is clearer in the residual velocity image than the original velocity map just above it.

\subsection{Clues to Orbital Dynamics in NGC 1365 - the NE Bridge and other Overshoot Regions}
\label{sec:expand}

An important capability of JWST is the ability to penetrate through the dust and identify structures that are not observable at optical wavelengths. 
An example 
is the NE Bridge, which is identified in the top panel of Figure~\ref{fig:reg_2}, but is much less obvious in the HST image in  Figure~\ref{fig:full_galaxy}.  Features like this bridge are often seen in hydro-dynamical simulations of bars (e.g., \citealt{sormani20}, \citealt{tress2020}, \citealt{henshaw2022}, Schinnerer et al. 2022, this volume). 


Gas in a barred potential travels almost radially along the bar dust lanes (in the frame co-rotating with the bar) towards the inner parts of the galaxy. 
However, not all the gas transitioning from the dust lanes to the inner regions does so smoothly - some overshoots and collides with gas travelling down the opposite dust lane. 

We believe the NE bridge feature is a clear example of material that has overshot the 
inner ring
after falling towards the center along the southern bar dust lane (bottom-right of Fig 1). 
When the material reaches the inner ring, gas and stars behave differently because they obey different equations of motion. Gas is affected by cloud collisions and thus dissipates some of its energy and eventually ends up accreting onto the inner ring. Stars, on the other hand, do not feel pressure forces and are unaffected by cloud collisions, so they fly undisturbed through the gas clouds and continue on their ballistic orbits towards the bar lane on the opposite (northern) bar lane, as appears to be the case with the NE Bridge.

These, and several other features obvious in the new JWST observations (e.g., the overshoot regions marked in  Figure \ref{fig:reg_1} and \ref{fig:reg_1_finding}), 
provide important insights into how the orbits of stars and gas vary depending on where they form, and how this results in the population of older clusters outside the inner star-formation ring.  A related phenomena  is discussed  in Whitmore et al. 2022, submitted (i.e., the ``300 My" overshoot region, as identified in Figure \ref{fig:reg_1_finding}). 

\section{Summary and Conclusions}
\label{sec:conclusions} 

In this work, we have presented new infrared JWST NIRCAM and MIRI imaging of the highly star-forming barred spiral galaxy NGC~1365 in the F200W, F300M, F335M, F360M, F770W, F1000W, F1130W, and F2100W filters.
These images show not only the same compact star clusters and individual stars observable in optical Hubble images, but also reveal recently formed clusters that remain deeply embedded in gas and dust, PAH emission, and new (likely dynamical) structures like the NE bridge and dark thin filaments.  From measurements of the concentration index 
we found that the F200W filter on JWST gives more accurate size measurement for stellar sources than the I filter on Hubble, resulting in a cleaner separation of stars and clusters.

We have compiled a comprehensive catalog of 37 massive $(M \gea 10^6~M_{\odot})$, young (Age $\lea 10$~My) clusters in NGC~1365, all but one located in the $28\arcsec \times 26\arcsec$ (2.7 kpc ×
2.5 kpc) region around the central 
star-forming ring.  We found that 4 of 30 ($\sim13$\%)  (i.e., including only the 30 objects with measured values below $10^6~M_{\odot}$ in Table \ref{tab:table_1}) are so deeply embedded that they cannot be easily seen in the optical
images taken by HST, even in the I filter, while 11 of 30 ($\sim37$\%) do not have mass
estimates from HST because they are missing in the HST B-band
images, and hence their ages cannot be easily determined in the visible using SED-fitting techniques.  These statistics imply that the  massive clusters in NGC~1365 remain deeply embedded (not detectable in the I-band) 
for $1.3\pm0.7$~My, and sufficiently embedded that they are not detectable in the B-band for $3.7\pm1.1$~My.
Hence, we find that NGC~1365 contains the richest known population of massive young clusters in any galaxy within 30~Mpc, even larger than the populations formed in the merging Antennae galaxy.

We also identified several interesting new features near the intersection between the bar and streamers of gas and dust that appear  to be colliding and triggering star formation.  Other new features include a NE bridge, which may be created by material that dynamically overshot the star-forming ring after falling towards the inner regions along the dust lane in the southern bar.

 The age, mass, and extinction estimates in this paper are approximate and preliminary, since they are based on correlations between optical  estimates and JWST observations for the more embedded clusters. In the future we expect to obtain better age estimates using SED fitting techniques using both the HST and JWST observations.   

\bigskip

\section*{Acknowledgements}
This work is based on observations made with the NASA/ESA/CSA JWST and Hubble Space Telescopes. The data were obtained from the Mikulski Archive for Space Telescopes at the Space Telescope Science Institute, which is operated by the Association of Universities for Research in Astronomy, Inc., under NASA contract NAS 5-03127 for JWST and NASA contract NAS 5-26555 for HST. The JWST observations are associated with program 2107, and those from HST with program 15454.  

\vspace{5mm}
\facilities{{\em HST} (Hubble Space Telescope), {\em JWST}, ALMA (Atacama Large Millimeter/submillimeter Array),  VLT-MUSE (Very Large Telescope - Multi Unit Spectroscopic Explorer).}

\vspace{5mm}

We would like to thank the referee for several useful and constructive comments that lead to improvements in  the paper. 
ATB would like to acknowledge funding from the European Research Council (ERC) under the European Union’s Horizon 2020 research and innovation programme (grant agreement No.726384/Empire).
FB would like to acknowledge funding from the European Research Council (ERC) under the European Union’s Horizon 2020 research and innovation programme (grant agreement No.726384/Empire)
JMDK gratefully acknowledges funding from the European Research Council (ERC) under the European Union's Horizon 2020 research and innovation programme via the ERC Starting Grant MUSTANG (grant agreement number 714907). COOL Research DAO is a Decentralized Autonomous Organization supporting research in astrophysics aimed at uncovering our cosmic origins.
MC gratefully acknowledges funding from the DFG through an Emmy Noether Research Group (grant number CH2137/1-1).
TGW acknowledges funding from the European Research Council (ERC) under the European Union’s Horizon 2020 research and innovation programme (grant agreement No. 694343).
This research was supported by the Excellence Cluster ORIGINS which is funded by the Deutsche Forschungsgemeinschaft (DFG, German Research Foundation) under Germany's Excellence Strategy - EXC-2094-390783311. Some of the simulations in this paper have been carried out on the computing facilities of the Computational Center for Particle and Astrophysics (C2PAP). We are grateful for the support by Alexey Krukau and Margarita Petkova through C2PAP.

This paper makes use of the following ALMA data:1447
ADS/JAO.ALMA\#2013.1.01161.S.
Based on observations collected at the European1459
Southern Observatory under ESO programmes 1100.B-1460
0651 (PHANGS-MUSE; PI: Schinnerer), as well as1461
094.B-0321 (MAGNUM; PI: Marconi).

\section*{Data Availability}

The imaging observations underlying this article can be retrieved from the Mikulski Archive for Space Telescopes at \url{https://archive.stsci.edu/hst/search_retrieve.html} under proposal GO-15654. High level science products, including science ready mosaicked imaging, associated with HST GO-15654 are provided at \url{https://archive.stsci.edu/hlsp/phangs-hst}.
The specific PHANGS-JWST observations analyzed can be accessed via \dataset[10.17909/9bdf-jn24]{http://dx.doi.org/10.17909/9bdf-jn24} and PHANGS-HST images accessed via \dataset[10.17909/t9-r08f-dq31]{https://dx.doi.org/10.17909/t9-r08f-dq31}.


\bibliography{all_nov_24_2021,phangsjwst}{}
\bibliographystyle{aasjournal}

\begin{table*}
 \caption{Census of Massive Young Star Clusters in NGC~1365}
 \label{tab:table_1}
 
     \centering

\begin{tabular}{lllllrrrllrr}
  \hline
 ID & Cl ID$^a$ & Other ID & RA & DEC & log Age$^b$ & log Mass$^c$ & Av$^d$ & CI$_{200}^e$ & Class$^f$ & Detected$^g$ &  Detected$^h$ \\
 
  & & & (J2000) & (J2000) & (H, H-JW) & (H, JW) & (H, JW) & & & B, V, I & PAHs \\
  & & & (deg) & (deg) & (Myr) &  (M$_\odot$) & (mag) & (mag) &  &  & \\

\hline  
 
   1    &       ---  &       ---     &       53.400281 &       -36.142577  &       ---,   6.5  &       ---, 6.2  &       ---,  12.5   &       2.05  &       3    &  ---, ---, I  & 3.3, 7.7, 11.3\\

  2    &       996   &       ---     &       53.399891  &       -36.142575  &       6.0, 6.0  &       7.1,     6.6  &       4.6,               4.4   &       1.83 &       2      &        B, V, I & 3.3, 7.7, 11.3 \\

    3    &       1051  &    J$^i$         &       53.400691  &       -36.142127  &       6.0, 6.0     &       7.2,     6.9  &       4.6,               5.4   &       1.70 &       2      &  B, V, I  & 3.3, 7.7, 11.3  \\  
    
  4    &       1064  &     ---         &       53.399386 &       -36.142085  &       6.6, 6.6    &       6.1     6.2  &       3.1, 4.3   &       1.69  &       2      &      B, V, I & 3.3, 7.7, 11.3 \\
  
  5    &       ---  &       ---     &       53.400338  &       -36.142072   &       ---,   6.5 &       ---,    6.00  &       ---,  4.9  &       1.64   &       2    &        B, V, I & 3.3, 7.7, 11.3 \\
  
  
  6    &       ---  &       ---     &       53.399056  &       -36.141991 &       ---,   6.5  &       ---,    5.8  &       ---, ---                  &       1.53 &       3    &       ---, ---, ---  & 3.3, 7.7, 11.3\\
  
  7    &       1116  &        M2$^j$, SSC3$^k$     &       53.399870  &       -36.141833 &       6.5,    6.5  &       6.9,     6.6  &       1.7,   1.0  &       1.45&       1      &       B, V, I & 3.3, 7.7, 11.3 \\ 
  
  8    &       1141  &       ---       &       53.401402  &       -36.141753  &       6.7,     6.7 &       6.6,     6.5  &       5.4,  4.6   &       2.05 &       2      &        B, V, I & 3.3, 7.7, 11.3 \\
  
  9    &       1251  &  ---            &       53.400184  &       -36.141315  &       6.0,     6.0  &       6.9,     6.6  &       4.6,               6.3   &       1.88 &       2      &        B, V, I & 3.3, 7.7, 11.3 \\
  
  10   &       1259  &       ---       &       53.402834 &       -36.141274  &       6.7,     6.7  &       6.0,     6.2   &       4.9,  5.8   &       1.59  &       2      &        B, V, I & 3.3, ---, --- \\

    11   &       --- &  ---     &       53.399771  &       -36.141127  &       ---,   6.5   &       ---.    6.5  &       ---, 4.5  &       1.66 &  2  &    B, V, I  &  3.3, 7.7, 11.3    \\

  12   &       1296  &   M3$^j$, SSC6$^k$         &       53.399552  &       -36.141121 &       6.5,    6.5 &       6.6,     6.6  &       1.2, 2.0  &       1.62  &       1      &     B, V, I  & 3.3, 7.7, 11.3   \\
  
  13   &    ---     &    ---        &       53.402935   &       -36.140825 &       ---,   6.5  &       ---,    5.9   &       ---,   ---                  &       1.39 &       3    &        ---, ---, ---  & 3.3, 7.7, --- \\
  
  14   &    ---     &       ---     &       53.402318 &       -36.140610  &       ---,   6.5  &       ---,    5.8  &       ---, 3.8  &       2.69 &       1    &      B, V, I  & ---,  7.7, 11.3 \\

  15   &       ---  &       ---     &       53.403212  &       -36.140610 &       ---,   6.5  &       ---,    6.3 &       ---, 12.2  &       1.52 &       3    &  ---, ---, I & 3.3, 7.7, ---       \\
  
  16   &       ---  &       ---     &       53.402911  &       -36.140455  &       ---,   6.5  &       ---,    5.8 &       ---, 8.0    &       1.73 &       2    &    B, V, I  & 3.3, 7.7, 11.3  \\
  
  17   &       ---  &       ---     &       53.402548 &       -36.140381   &       ---,   6.5 &       ---,    6.0  &       ---, ---                  &       1.79 &       3    &   ---, ---, ---   & 3.3, 7.7, 11.3    \\
  
  18   &       1506  &      ---        &       53.403394 &       -36.140211 &       6.0,     6.0  &       6.4     6.2  &       4.6,               6.5  &       1.76 &       2      &   B, V, I  & 3.3, ---, ---   \\
  
  19   &       ---  &       ---     &       53.399524  &       -36.140090 &       ---,   6.5 &       ---,    6.5  &       ---, 8.4   &       1.89 &       3    &       ---, V, I & 3.3, 7.7, 11.3 \\
  
  20   &       1528  &  ---            &       53.403379  &       -36.140083  &       6.5,    6.5  &       6.2,     6.1   &       4.6, 6.1   &       1.74 &       2      &     B, V, I  & 3.3, 7.7, --- \\
  
  21   &       ---  &       ---     &       53.403331  &       -36.139860  &       ---,   6.5  &       ---,    6.2  &       ---, 6.9  &       1.52 &       2    &     B, V, I  & 3.3, 7.7, 11.3  \\
  
  22   &       ---  &       ---     &       53.400461  &       -36.139849  &       ---,   6.5  &       ---,    6.7  &       ---, 7.4    &       1.72 &       2    &   B, V, I  & 3.3, 7.7, 11.3   \\
  
  23   &       ----  &       ---     &       53.403374  &       -36.139568 &       ---,   6.5  &       ---,    6.3 &       ---, 5.6   &       1.74  &       2    &     B, V, I  & 3.3, 7.7, --- \\
  
  
  24   &       1640  &      ---        &       53.399837   &       -36.139432   &       6.0,     6.0  &       6.3,     5.8  &       4.6, 2.4  &       2.03 &       1      &    B, V, I  & 3.3, ---, ---  \\
  
  25   &       ---  &       ---     &       53.400928  &       -36.139100  &       ---,   6.5 &       ---, 6.7  &       ---, ---                  &       1.84 &       3    &   ---, ---, I  & 3.3, ---, 11.3  \\

  26   &       1700  &   C1$^i$         &       53.404026 &       -36.139017  &       6.7,     6.7 &       6.9, 6.7  &       4.6,               2.9   &       1.75 &       1      &    B, V, I  & 3.3, 7.7, 11.3 \\
  
  27   &       ---  &       ---     &       53.402918  &       -36.138860  &       ---,   6.5  &       ---, 6.6  &       ---, 9.5    &       1.79 &       3    &    ---, ---, I    & 3.3, ---, ---  \\
  
  28   &       ---  &       M6$^j$, G$^i$      &       53.403181 &       -36.138475  &       ---,   6.5  &       ---, 6.6  &       ---,  19.3                  &       1.85 &       3      &    ---, ---, I  & 3.3, SAT, 11.3   \\
  
  29   &       ---  &       M4$^j$, D$^i$     &       53.401702  &       -36.138433  &       ---,   6.5  &       ---,    6.0  &       ---,   ---                  &       1.62 &       3      &    ---, ---, --- & 3.3, SAT, 11.3  \\
  
  30   &       1782  &  ---            &       53.400982  &       -36.138416  &       6.5,    6.5 &       6.2,     6.3  &       4.9,   7.5   &       1.55 &       2      &    B, V, I   & 3.3, 7.7, 11.3  \\
  
  31   &       1794  &      ---        &       53.401669  &       -36.138367  &       6.6,     6.6  &       6.1,   6.0  &       4.6,               5.8   &       1.68 &       2      &    B, V, I  & 3.3, 7.7, 11.3   \\
  
  32   &       1810  &  ---            &       53.401369  &       -36.138271  &       6.9,     6.9  &       6.2, 6.8  &       4.6,               7.7  &       1.97  &       2      &    ---, V, I  & 3.3, ---, ---\\
  
  33   &       1920  &   M5$^j$, E$^i$         &       53.402514  &       -36.137680  &       ---,   6.5 &       ---,    6.6  &       ---, 4.3   &       1.55  &       2      &    B, V, I   & 3.3, SAT , 11.3  \\
  
  34   &       1932  &      ---        &       53.402534  &       -36.137596  &       6.7,     6.7  &       6.4,     5.8   &       5.0, 7.4   &       1.45 &       2      &     B, V, I  & 3.3, 7.7, 11.3  \\
  
  35   &       1948  &  ---            &       53.403066 &       -36.137494  &       6.5    6.5  &       6.5,     5.8  &       6.9, 6.6   &       1.57 &       2      &    B, V, I   & 3.3, 7.7, 11.3   \\
  
  36   &       ---  &       ---     &       53.403115  &       -36.137151 &       ---,   6.5 &       ---,   6.3  &       ---, 6.4    &       1.42 &       2    &    B, V, I    & 3.3, 7.7, 11.3  \\
  
  37   &       2242  &      ---        &       53.407333  &       -36.135815  &       6.6,     6.5  &       6.2,     6.3  &       2.0, 2.6   &       1.71  &       1      &    B, V, I  & 3.3, 7.7, 11.3  \\

  \hline
  
 \hline
 \end{tabular}

 \raggedright
 
$^a$ - Cluster ID from the NGC 1365 compact cluster catalog, as found at https://archive.stsci.edu/hlsp/phangs-hst. \\
$^b$ - Log Age - First column: Hubble value from Whitmore et al. 2022 (submitted). Typical errors are 0.4 dex. Second column: Adopted value from first column if available (Hubble), or 6.5 otherwise (JWST - typical errors are 0.6 dex). See Section \ref{sec:ages_for_embedded} for discussion. \\
$^c$ - Log Mass - First column: Hubble value from Whitmore et al. 2022 (submitted). Second column: Value from mean of values based on F200W formula (i.e., $\mbox{log}~(M/M_{\odot}) = - 0.4068 \times m_{\rm F200W} + 13.83$) and F300M formulae (i.e., $\mbox{log}~(M/M_{\odot}) = - 0.3639 \times m_{\rm F300M} + 13.25$). Typical errors are 0.35 dex. See Section \ref{sec:mass}. \\
$^d$ - Extinction (A$v$) - First column: Hubble value from Whitmore et al. 2022 (submitted). Second column: Value from formula 
$A_V (\mbox{JWST) = [(m$_I$ - m$_{F360M}$)}  \times 0.810 + 0.99] \times 3.1$. Typical errors are 1.0 mag). See Section \ref{sec:extinction}. \\
$^e$ - Concentration Index (CI) using the difference in aperture magnitudes in the  F200W filter using radii of 1 and 4 pixels. See Section \ref{sec:size}.\\

\end{table*}

\clearpage
Notes from table (continued):

\bigskip 

\noindent $^f$ - Visual estimate of how embedded the cluster is. Class 1 = largely unobscurred (A$_V \lea$ 3 mag)  , Class 2 = partly embedded (3 $\lea A_V \lea$ 10 mag), Class 3 = embedded (A$_V \gea$ 10 mag).\\
$^g$ - Brighter than 26th mag in relevant band.  \\
$^h$ - Brighter than 20th mag in relevant band. \\
$^i$ - Radio Source ID - \citep{sandqvist95}. \\
$^j$ - MID IR ID - (\citealt{galliano08}).\\
$^k$ - SSC = Super Star Cluster ID   - \citep{kristen97}.\\

\end{document}